\documentclass[final]{aipproc}

\usepackage{amsmath,amssymb,graphicx,bm}

\newcommand{\be}{\begin{equation}}
\newcommand{\ee}{\end{equation}}
\newcommand{\lm}{\Lambda}
\newcommand{\vlowk}{V_{{\rm low}\,k}}

\newcommand{\fmi}{\, \text{fm}^{-1}}

\newcommand{\mev}{\, \text{MeV}}

\newcommand{\ohfnn}{\Omega_{1,{\rm NN}}}
\newcommand{\ohfnnn}{\Omega_{1,{\rm 3N}}}
\newcommand{\oa}{\Omega_{2,{\rm a}}}
\newcommand{\on}{\Omega_{2,{\rm n}}}

\layoutstyle{6x9}

\begin{document}

\title{The neutron matter equation of state from low-momentum interactions}


\author{L. Tol\'os}{
address={Gesellschaft f\"ur Schwerionenforschung,
Planckstrasse 1, D-64291 Darmstadt, Germany}}
\author{B. Friman}{
address={Gesellschaft f\"ur Schwerionenforschung,
Planckstrasse 1, D-64291 Darmstadt, Germany}}
\author{A. Schwenk}{
address={TRIUMF, 4004 Wesbrook Mall, Vancouver, BC, Canada, V6T 2A3},
altaddress={Department of Physics, University of Washington, Seattle, 
WA 98195-1560}}

\begin{abstract}
We calculate the neutron matter equation of state at finite temperature 
based on low-momentum nucleon-nucleon and three-nucleon interactions.
Our results are compared to the model-independent virial equation of 
state and to variational calculations. We provide a simple estimate for 
the theoretical error, important for extrapolations to astrophysical 
conditions.
\end{abstract}

\maketitle

The nuclear equation of state plays a central role in the physics of 
neutron stars~\cite{LP} and core-collapse supernovae~\cite{Mezza,Janka}.
Renormalization group methods coupled with effective field theory
offer the possibility of a new and systematic approach to nuclear 
matter: For low-momentum interactions $\vlowk$~\cite{VlowkReport} 
with cutoffs around $2 \fmi$, the strong short-range repulsion in
conventional nucleon-nucleon (NN) interactions and the tensor force 
are tamed~\cite{Vlowknucmatt,VlowkWeinberg}. At sufficient density, 
Pauli blocking eliminates the shallow bound states, and thus the 
particle-particle channel becomes perturbative in nuclear 
matter~\cite{Vlowknucmatt}. In addition, the corresponding 
leading-order chiral three-nucleon (3N) interaction becomes 
perturbative in light nuclei for $\lm \lesssim 2 \fmi$~\cite{Vlowk3N}. 
Consequently, the Hartree-Fock (HF) approximation is a good starting point,
and perturbation theory (in the sense of a loop expansion) around the 
HF energy becomes tractable~\cite{Vlowknucmatt}. The perturbative
character is due to a combination of Pauli blocking and an appreciable 
effective range (see also~\cite{dEFT}).

At finite temperature, the loop expansion around the HF free
energy can be realized, based on the work of Kohn, Luttinger
and Ward~\cite{KL,LW}, by the perturbative expansion of the free
energy, where the momentum dependence of the self-energy is
treated 
perturbatively. In this work, we include the first-order NN and 3N 
contributions, as well as anomalous and normal second-order 
diagrams with NN interactions. The pressure, entropy and
energy are calculated using standard thermodynamic 
relations. We use the cutoff dependence to provide simple
error estimates, and find that the cutoff dependence is
reduced significantly, when second-order contributions are
included.

We start from the perturbative expansion of the grand-canonical
potential $\Omega(\mu,T,V)$, where $\mu$ is the chemical potential,
$T$ the temperature and $V$ the volume:
\be
\Omega = \Omega_0 +  \ohfnn + \ohfnnn  +  \oa + \on  + \ldots \,.
\ee
The non-interacting system is given by $\Omega_0$, $\Omega_1 
= \ohfnn + \ohfnnn$ denotes the first-order NN and 3N, and 
$\oa + \on$  are the second-order anomalous and normal 
contributions.

The free energy $F(N,T,V)$ 
is obtained by a Legendre transformation of the 
grand-canonical potential with respect to the chemical potential,
$F(N,T,V) = \Omega(\mu,T,V) +\mu N$, with mean particle number $N$. 
Following Kohn and Luttinger~\cite{KL}, we have
\be
F(N) = F_0(N) + \Omega_1(\mu_0)+\on(\mu_0) +
\left[ \oa(\mu_0) - \frac{1}{2} \, \frac{(\partial \ohfnn/\partial
\mu)^2}{\partial^2 \Omega_0/\partial \mu^2} \biggr|_{\mu_0}
\right] + \ldots \,,
\label{eq:F}
\ee
where $\mu_0$ is the chemical potential of a non-interacting system
with the same density $\rho = N/V$ as the interacting system, $N = 
- [\partial \Omega_0/\partial \mu]_{\mu_0}$, and $F_0(N) = 
\Omega_0(\mu_0) + \mu_0 N$ is the free energy of the non-interacting 
system. The above expansion ensures that the $T \to 0$ limit is
correctly reproduced~\cite{KL,LW}. The anomalous second-order diagram 
accounts for perturbative corrections to the free single-particle spectrum.

\begin{figure}
\includegraphics[clip=,width=4.1in]{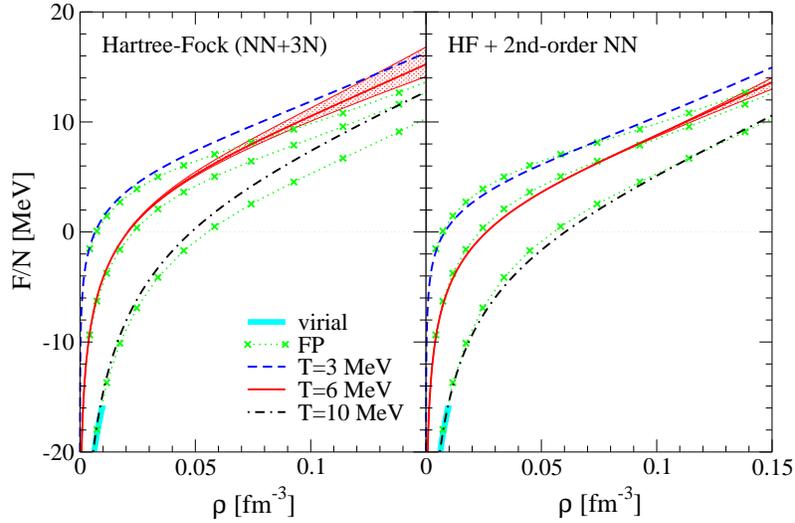}
\caption{The free energy per particle
$F/N$ as a function of density $\rho$. The
left figure gives the first-order NN and 3N contributions with a free 
single-particle spectrum. Second-order anomalous and normal
NN contributions are included 
in the right figure. Our results are compared to the virial equation 
of state (virial)~\cite{vEOSneut} and to the variational calculations 
of Friedman and Pandharipande (FP)~\cite{FP}. The virial curve ends
where the fugacity $z=e^{\mu/T}$ is $0.5$.}
\label{free_energy}
\end{figure}

Our results~\cite{tolos} 
for the free energy per particle are shown in Fig.~\ref{free_energy}
for temperatures $T=3 \mev, 6 \mev$ and $10 \mev$,
where the low-momentum interaction $\vlowk$ is obtained from the
Argonne $v_{18}$ potential for a cutoff $\lm = 2.1 \fmi$. For the
3N contribution at the HF level, 
we find that only the $c_1$ and $c_3$ terms of 
the long-range $2\pi$-exchange part survive (for details on the 
3N interaction, see~\cite{Vlowknucmatt,Vlowk3N}). For the $T=6 \mev$ 
results, we provide error estimates by varying the cutoff over 
the range $\lm=1.9 \fmi$ (lower curve) to $\lm=2.5 \fmi$ (upper curve).
As expected the error grows with increasing density. From 
Fig.~\ref{free_energy}, we observe that the equation of state
becomes significantly less cutoff dependent with the inclusion 
of the second-order NN contributions.

In Fig.~\ref{free_energy}, we also compare our results for the 
free energy to the
model-independent virial equation of state~\cite{vEOSneut} and
to the variational calculations of Friedman and 
Pandharipande~\cite{FP} (FP,
based on the Argonne $v_{14}$ and a 3N potential). 
We find a very good agreement with the virial free energy, and
for the densities in Fig.~\ref{free_energy} similar results as FP.

\begin{figure}
\includegraphics[clip=,width=4.1in]{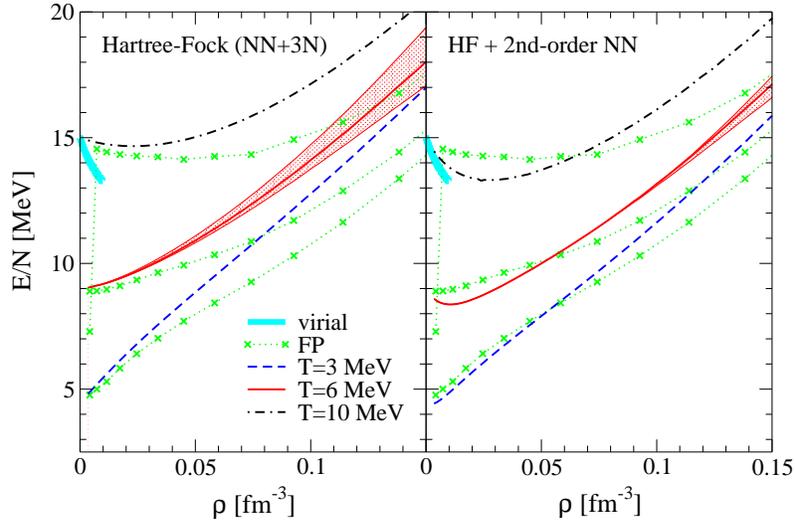}
\caption{The energy per particle
$E/N$ as a function of density $\rho$ to first
and second order as in Fig.~\ref{free_energy}.}
\label{energy}
\end{figure}

Our results~\cite{tolos} for the energy per particle are presented
in Fig.~\ref{energy}. As for the free energy, we observe additional 
binding and a significantly reduced cutoff dependence at second 
order. In contrast to the variational calculation of FP~\cite{FP},
the low-density behavior at second order is in good agreement with 
the virial equation of state~\cite{vEOSneut}. This highlights the 
importance of a correct finite-temperature treatment of second
and higher-order contributions. This work is part of a program 
to improve the nuclear equation of state input for astrophysics,
and to provide error estimates, for example, for the neutron star 
mass and radius predictions.

\vspace*{-2mm}

\begin{theacknowledgments}
This work was supported in part by the Virtual Institue VH-VI-041
of the Helmholtz Association, NSERC and the US DOE Grant
DE--FG02--97ER41014. TRIUMF receives federal funding via a
contribution agreement through the NRC.
\end{theacknowledgments}


\vspace*{-2mm}

\begin{thebibliography}{9}
\bibitem{LP} J.M.\ Lattimer and M.\ Prakash, \emph{Astrophys.\ J.}
\textbf{550}, 426 (2001).

\bibitem{Mezza} A.\ Mezzacappa, \emph{Annu.\ Rev.\ Nucl.\ Part.\ Sci.} 
\textbf{55}, 467 (2005).

\bibitem{Janka} H.T.\ Janka, R.\ Buras, F.S.\ Kitaura Joyanes, A.\ Marek
and M.\ Rampp, astro-ph/0405289.

\bibitem{VlowkReport} S.K.\ Bogner, T.T.S.\ Kuo and A.\ Schwenk,
\emph{Phys.\ Rept.} \textbf{386}, 1 (2003).

\bibitem{Vlowknucmatt} S.K.\ Bogner, A.\ Schwenk, R.J.\ Furnstahl and
A.\ Nogga, \emph{Nucl.\ Phys.} \textbf{A763}, 59 (2005).

\bibitem{VlowkWeinberg} S.K.\ Bogner, R.J.\ Furnstahl, S.\ Ramanan and
A.\ Schwenk, \emph{Nucl.\ Phys.} \textbf{A773}, 203 (2006).

\bibitem{Vlowk3N} A.\ Nogga, S.K.\ Bogner and A.\ Schwenk, \emph{Phys.\ Rev.}
\textbf{C70}, 061002(R) (2004).

\bibitem{dEFT} A.\ Schwenk and C.J.\ Pethick, \emph{Phys.\ Rev.\ Lett.} 
\textbf{95}, 160401 (2005).

\bibitem{KL} W.\ Kohn and J.M.\ Luttinger, \emph{Phys.\ Rev.} \textbf{118},
41 (1960).

\bibitem{LW} J.M.\ Luttinger and J.C.\ Ward, \emph{Phys.\ Rev.} \textbf{118},
1417 (1960).

\bibitem{tolos} L.\ Tol\'os, B.\ Friman and A.\ Schwenk, in preparation.

\bibitem{vEOSneut} C. \ J. \ Horowitz and A.\ Schwenk, \emph{Phys. \ Lett.} \textbf{B638}, 153 (2006).

\bibitem{FP} B. \ Friedman and V. \ R. \ Pandharipande, \emph{Nucl.\ Phys.} \textbf{A361}, 502 (1981).

\end{thebibliography}
\end{document}